\documentstyle[aps,psfig,preprint]{revtex}
\tightenlines
\begin{document}
\title{Parameters estimation in quantum optics}
\author{G. Mauro D'Ariano, Matteo G. A. Paris and Massimiliano F. Sacchi}
\address{Theoretical Quantum Optics Group\\ 
Unit\`a INFM and Dipartimento di
fisica ``Alessandro Volta''\\Universit\`a di Pavia, via A. Bassi 6, I-27100 
Pavia, ITALY} \date{\today} \maketitle
\begin{abstract}
We address several estimation problems in quantum optics by means of
the maximum-likelihood principle. We consider Gaussian state estimation 
and the determination of the coupling parameters of quadratic Hamiltonians.  
Moreover, we analyze different schemes of phase-shift estimation. Finally, 
the absolute estimation of the quantum efficiency of both linear and 
avalanche photodetectors is studied. In all the considered applications, 
the Gaussian bound on statistical errors is attained with a few
thousand data.
\end{abstract}
\section{Introduction}\label{s:intro}
In order to gain information about a physical quantity one should, in
principle, measure the corresponding quantum observable. In cases when 
the measurement can be directly implemented the statistics of the outcomes
is governed (in ideal conditions, i.e. neglecting thermal, mechanical
or other sources of classical noise) only by the intrinsic fluctuations 
of the observable, namely by the quantum nature of the system under 
investigation. In practice, however, it is most likely that the desired 
observable does not correspond to a feasible measurement scheme, or the 
physical quantity does not correspond to any observable at all. 
In such case one has to {\em infer} the value of the quantity of 
interest from the measurement of a different observable, or generally 
of a set of observables. In this situation, even in ideal conditions, 
the indirect parameter estimation gives an additional uncertainty for 
the estimated value, and the quantum estimation theory \cite{hel,hol} 
provides a general framework to optimize the inference procedure. \par 
In the recent years, the indirect reconstruction of observables and 
quantum states has received much attention. Among the many reconstruction 
techniques, the most successful is quantum homodyne tomography \cite{qht}, 
which, indeed, is the only method which has been experimentally implemented
\cite{nat}.  Quantum tomography provides the complete characterization
of the state, i.e.  the reconstruction of any quantity of interest by
simple averages over experimental data.  In many cases, however, one
may be interested not in the complete characterization of the state,
but only in some specific feature, like the phase or the amplitude of
the field.  Moreover, one can address the problem of characterizing an
optical device, rather than a quantum state, like measuring the
coupling constant of an active medium or the quantum efficiency of a
photodetector. In all these cases, the desired parameter does not
correspond to a measurable observable, and contains only a partial 
information about the quantum state of light involved in the process. 
Our goal is to link the estimation of such parameters with the results 
from feasible measurement schemes, as homodyne, heterodyne or direct
detection, and to make the   estimation procedure the most efficient. 
\par Among all possible procedures for parameter estimation, the
maximum-likelihood (ML) method is, in the sense discussed below,
the most general, and widely usable in practice. The ML procedure
answers to the following question: which values of the parameters are
most likely to produce the results which we actually observe in
the measurement~?  This statement can be quantified, and the resulting 
procedure is the ML estimation of the parameters. \par 
Recently, the ML principle has been applied to the reconstruction of 
the whole state of a generic quantum system \cite{kon,hra}. In that 
case the parameters of interest are the matrix elements of the density 
operator in a suitable representation. Bayesian and ML approaches have 
been also applied in neutron interferometry \cite{rau}. \par 
In this paper, we focus our attention on the determination of 
specific parameters which are relevant in quantum optics, and analyze 
their ML estimation procedure in some details. \par 
In the next Section we briefly review the ML estimation procedure 
as well as the method to evaluate its precision. In Section \ref{s:gauss} 
we consider the estimation of the parameters of a Gaussian state and 
of the coupling constants of a generic quadratic single-mode Hamiltonian.  
As we will show, the two estimation problems are closely related, and 
ML principle leads to a fully general solution. In Section \ref{s:phase} 
we study different schemes of phase estimation, whereas in Section 
\ref{s:eta} the ML principle is applied to the estimation of the 
quantum efficiency of both linear and avalanche photodetectors.  
Section \ref{s:outro} closes the paper by summarizing our results.
\section{Maximum-likelihood estimation}
Here we briefly review the theory of the maximum-likelihood (ML)
estimation of a single parameter. The generalization to several
parameters is straightforward.  Let $p(x | \lambda)$ the probability
density of a random variable $x$, conditioned to the value of the
parameter $\lambda$.  The form of $p$ is known, but the true value of
the parameter $\lambda$ is unknown, and will be estimated from the
result of a measurement of $x$.  Let $x_1, x_2, ..., x_N$ be a random
sample of size $N$. The joint probability density of the independent
random variable $x_1, x_2, ..., x_N$ (the global probability of the
sample) is given by
\begin{eqnarray}
{\cal L}(x_1, x_2, ..., x_N| \lambda)= \Pi_{k=1}^N \: p(x_k |\lambda)
\label{likdef}\;,
\end{eqnarray}
and is called the likelihood function of the given data sample
(hereafter we will suppress the dependence of ${\cal L}$ on the data). The
maximum-likelihood estimator $\lambda_{\sc ml} \equiv \lambda_{\sc ml} 
(\{x_k\})$ (MLE) of the parameter $\lambda$ maximizes ${\cal L} 
(\lambda)$ for variations of $\lambda$, namely it is given by the 
solution of the equations
\begin{eqnarray}
\frac{\partial {\cal L} (\lambda) } {\partial \lambda} = 0 \; ; \quad 
\frac{\partial^2 {\cal L} (\lambda)} {\partial \lambda^2} < 0
\label{maxlikdef}\;.
\end{eqnarray}
Since the likelihood function is positive the first equation is
equivalent to
\begin{eqnarray}
\frac{\partial L (\lambda)}{\partial \lambda} = 0 \; , 
\label{loglikdef}\;
\end{eqnarray}
where 
\begin{eqnarray}
L(\lambda) = \log {\cal L} (\lambda) = \sum_{k=1}^N \log p(x_k | \lambda)
\label{loglikfun}\;
\end{eqnarray}
is the so-called log-likelihood function. 
The form of the ML principle in Eq. (\ref{loglikdef}) is the most 
often used in practice.  
\par The importance of MLE stems from the following 
theorems \cite{cramer,kendall}
\begin{enumerate}
\item Maximum-likelihood estimators are consistent, i.e. they converge
in probability to the true value of the parameter for increasing size of
the data sample. 
\item The distribution of MLEs tends to the normal distribution in the
limit of large samples, and MLEs have minimum variance. For finite
samples the variance is governed by the Cram\'er-Rao bound (see below).
\end{enumerate}
There are also situations in which the MLE gives a poor estimation for
a parameter. However, for the distributions considered here 
the ML procedure is statistically efficient. \par
In order to obtain a measure for the confidence interval in  
the determination of $\lambda_{\sc ml}$ we consider the variance
\begin{eqnarray}
\sigma^2_\lambda = \int \left[\prod_k dx_k \: p(x_k|\lambda)\right] 
\left[\lambda_{\sc ml} 
(\{x_k\})- \lambda \right]^2 \: \label{varMLEdef}\;.
\end{eqnarray}
Upon defining the Fisher information 
\begin{eqnarray}
F= \int dx \left[ \frac{\partial p(x |\lambda)}{\partial \lambda}\right]^2
\frac1{p(x | \lambda)}
\label{FisherDef}\;,
\end{eqnarray}
it is easy to prove \cite{tucker} that 
\begin{eqnarray}
\sigma^2_\lambda  \geq \frac1{N\:F}  \label{CRFisher}\;,
\end{eqnarray}
where $N$ is the number of measurements. The inequality in Eq. 
(\ref{CRFisher}) is known as the Cram\'er-Rao bound \cite{cramer}
on the precision of ML estimation. Notice that this bound holds 
for any functional form of the probability distribution 
$p(x|\lambda)$, provided that the Fisher information exists $\forall \lambda$
and $\partial_\lambda p(x|\lambda)$ exists $\forall x$. When an experiment 
has "good statistics" (i.e. a data sample large enough) the Cram\'er-Rao 
bound is saturated. As we will show in the following, the application 
of the ML principle in quantum optics generally corresponds to 
estimators for which the Cram\'er-Rao bound is attained with a relatively 
small number of measurements, i.e. the ML procedure provides an efficient 
estimation of the parameters. In Sections \ref{s:gauss} and \ref{s:phase}
examples will be examined where the probability $p(x|\lambda)$ is Gaussian
versus $x$ and not Gaussian versus the parameter $\lambda$, whereas in 
Section \ref{s:eta} an example with discrete measurement outcomes
($x=0,1$) will be also analyzed.
\section{Gaussian-state estimation}\label{s:gauss} 
In this section we apply the ML method to estimate the
quantum state of a single-mode radiation field that is characterized by a
Gaussian Wigner function. Such kind 
of states represents the wide class of coherent, squeezed and thermal
states. Apart from an irrelevant phase, we consider the Wigner
function of the form 
\begin{eqnarray}
W(x,y)=\frac{2\Delta ^2}{\pi}\exp\left\{-2\Delta ^2\left[
e^{-2r}(x-\hbox{Re}\mu)^2
+e^{2r}(y-\hbox{Im}\mu)^2\right]\right\}\;,\label{wxy}
\end{eqnarray}
and we apply the ML technique with homodyne detection 
to estimate the four real parameters 
$\Delta , r, \hbox{Re}\mu$ and $\hbox{Im}\mu$. 
The four parameters provide the number of thermal, squeezing and 
coherent-signal photons in the quantum state as follows
\begin{eqnarray}
& & n_{th}=\frac 12\left(\frac {1}{\Delta ^2}-1\right)\;,
\nonumber \\
&&n_{sq}=\sinh^2 r\;, \nonumber \\& & n_{coh}=|\mu|^2\;.
\end{eqnarray}
In terms of density matrix, the state corresponding to the Wigner
function in Eq. (\ref{wxy}) writes
\begin{eqnarray}
\varrho =D(\mu)\,S(r)\,\frac
{1}{n_{th}+1}\left(\frac{n_{th}}{n_{th}+1}\right)^{a^\dag a}
\,S^\dag (r)\,D^\dag (\mu)\;,
\end{eqnarray}
where $S(r)=\exp[r(a^2-a^{\dag 2})/2]$ and $D(\mu)=\exp(\mu a^\dag -\mu
^*a)$ denote the squeezing and displacement operators, respectively. 
\par The theoretical homodyne probability distribution at phase $\phi
$ with respect to the local oscillator is given by the Gaussian \cite{yuen} 
\begin{eqnarray}
p(x,\phi)&=&\sqrt{\frac{2 \Delta ^2}{\pi(e^{2r}\cos^2\phi
+e^{-2r}\sin^2\phi)}}\nonumber \\&\times &  
\exp\left\{-\frac{2 \Delta ^2}{e^{2r}\cos^2\phi
+e^{-2r}\sin^2\phi}\left[x-\hbox{Re}(\mu\,e^{-i\phi})\right]^2\right\}
\;.\label{pxfi}
\end{eqnarray}
From Eqs. (\ref{loglikfun}) and (\ref{pxfi}) one easily evaluates 
the log-likelihood
function for a set of $N$ homodyne outcomes $x_i$ at random phase
$\phi _i$ as follows
\begin{eqnarray}
L&=&\sum_{i=1}^N \frac 12 \log \frac{2\Delta^2}{\pi 
(e^{2r}\cos^2 \phi_i
+e^{-2r}\sin^2\phi_i)} \nonumber \\&- & 
\frac{2 \Delta ^2}{e^{2r}\cos^2\phi_i
+e^{-2r}\sin^2\phi_i}\left[x_i-\hbox{Re}(\mu\,e^{-i\phi_i})\right]^2
\;.\label{lgau}
\end{eqnarray}
The ML estimators $\Delta _{\sc ml}, r_{\sc ml}, \hbox{Re}\mu _{\sc
ml}$ and $\hbox{Im}\mu_{\sc ml}$ are found upon maximizing
Eq. (\ref{lgau}) versus $\Delta, r, \hbox{Re}\mu$ and $\hbox{Im}\mu$.
\par
\begin{figure}[h]
\begin{center}
\centerline{\psfig{file=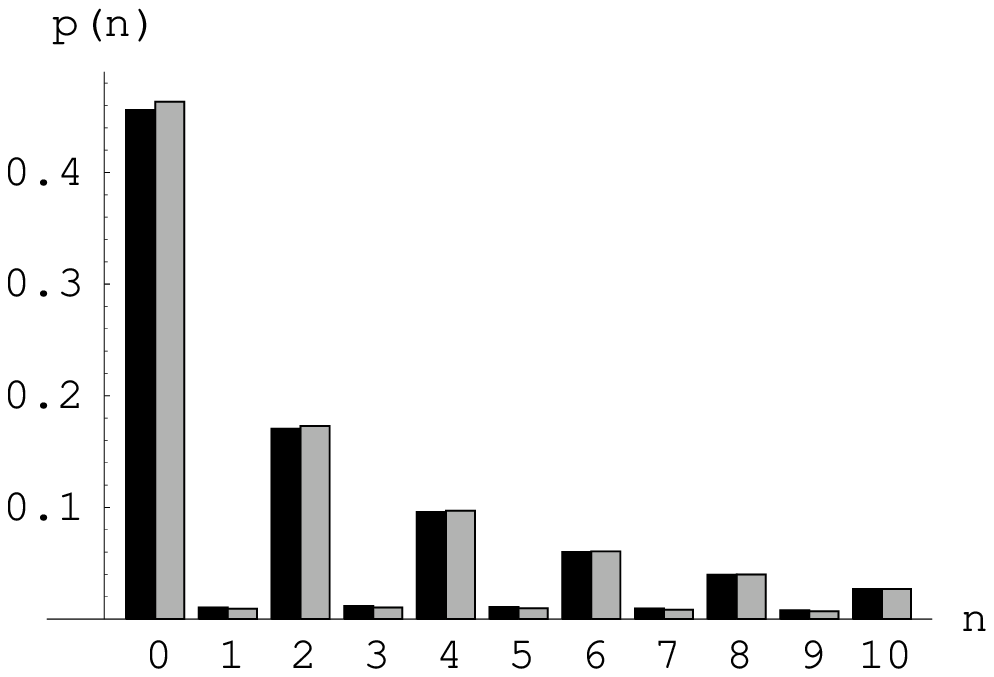,width=8cm}}
\end{center} 
\caption{Photon-number probability of a squeezed-thermal state. 
The black histogram for the theoretical, the gray one 
reconstructed state by means of the maximum likelihood method and homodyne
detection. Number of data samples $N=50000$, quantum
efficiency $\eta =80\%$, number of thermal photons $n_{th}=0.1$,
number of squeezing photons $n_{sq}=3$. The statistical error affects
the third decimal digit, and  it is not visible in the scale of the
plot.}
\label{f:sqth}
\end{figure}

 In order to obtain a global estimation of the goodness of the state
reconstruction, we evaluated the normalized overlap $\cal O$ between
the theoretical and the estimated state
\begin{eqnarray}
{\cal O}=\frac{\hbox{Tr}[\varrho \,\varrho _{\sc ml}]}{\sqrt
{\hbox{Tr}[\varrho ^2]\,\hbox{Tr}[\varrho _{\sc ml} ^2]}}\;.
\end{eqnarray}
Notice that ${\cal O}=1$ iff $\varrho =\varrho _{\sc ml}$. Through
some Monte-Carlo simulations, we always found a value around unity,
typically with statistical fluctuations over the third digit, for
number of data samples $N=50000$, quantum efficiency at homodyne
detectors $\eta=80\%$, and state parameters with the following ranges:
$n_{th}<3$, $n_{coh}<5$, and $n_{sq}<3$. Also with such a small number
of data samples, the quality of the state reconstruction is so good
that other physical quantities that are theoretically evaluated from
the experimental values of $\Delta _{\sc ml}, r_{\sc ml}, \hbox{Re}\mu
_{\sc ml}$ and $\hbox{Im}\mu_{\sc ml}$ are inferred very precisely. 
For example, we evaluated the photon number probability of a squeezed thermal
state, which is given by the integral
\begin{eqnarray}
\langle n|\varrho |n\rangle =\int _{0}^{2\pi}
\frac {d\phi}{2\pi} \frac{[C(\phi,n_{th},r)-1]^n}
{C(\phi,n_{th},r)^{n+1}}\;,
\end{eqnarray}
with $C(\phi,n_{th},r)=(n_{th}+\frac
12)(e^{-2r}\sin^2\phi+e^{2r}\cos^2\phi)+\frac 12$.  The comparison
of the theoretical and the experimental results for a state with
$n_{th}=0.1$ and $n_{sq}=3$ is reported in Fig. \ref{f:sqth}. The
statistical error of the reconstructed number probability affects the
third decimal digit, and is not visible on the scale of the plot.
\par The estimation of parameters of  Gaussian Wigner functions through
the ML method allows one to estimate the parameters in quadratic
Hamiltonians of the generic form
\begin{eqnarray}
H=\alpha a+\alpha ^* a^\dag + \varphi a^\dag a +\frac 12 \xi a^2+\frac
12 \xi^*a^{\dag 2}\;.\label{ham}
\end{eqnarray}
In fact, the unitary evolution operator $U=e^{-iHt}$ preserves the
Gaussian form of an input state with Gaussian Wigner function. In
other words, one can use a Gaussian state to probe and characterize an
optical device described by a Hamiltonian as in Eq. (\ref{ham}).  
Assuming $t=1$ without loss of generality, 
the Heisenberg evolution of the radiation mode $a$ is given by
\begin{eqnarray}
U^\dag \,a\,U=\gamma a+\delta a^\dag +\mu\;,
\end{eqnarray}
with 
\begin{eqnarray}
&&\gamma =\cos (\sqrt{\varphi ^2-|\xi|^2})-i\frac {\varphi}
{\sqrt{\varphi ^2-|\xi|^2}}\sin (\sqrt{\varphi ^2-|\xi|^2})\;,
\nonumber \\& & \delta =-i \frac{\xi ^*}{\sqrt{\varphi ^2-|\xi|^2}}
\sin(\sqrt{\varphi ^2-|\xi|^2})\;, \nonumber \\& & 
\mu=\frac{\varphi\alpha ^*-\xi^*\alpha }{\varphi ^2-|\xi|^2}
(\cos (\sqrt{\varphi ^2-|\xi|^2})-1)-i\frac{\alpha ^*}
{\sqrt{\varphi ^2-|\xi|^2}}\sin(\sqrt{\varphi ^2-|\xi|^2})\;.\label{3eq}
\end{eqnarray}
For an input state $\varrho $ 
with known Wigner function $W_\varrho (\beta \,,\beta ^*)$, the
corresponding output Wigner function writes
\begin{eqnarray}
W_{U\varrho U^\dag }(\beta \,,\beta ^*)=
W_\varrho [(\beta -\mu)\gamma ^*-(\beta ^*-\mu ^*)\delta \,,
(\beta^* -\mu^*)\gamma -(\beta -\mu )\delta ^*]\;.\label{wout}
\end{eqnarray}
Hence, by estimating the parameters $\gamma ,\delta ,\mu $ and
inverting Eqs. (\ref{3eq}), one obtains the ML values for $\alpha
,\varphi $, and $\xi $ of the Hamiltonian in Eq. (\ref{ham}). 
The present example can be used in practical applications for the
estimation of the gain of a phase-sensitive amplifier or equivalently
to estimate a squeezing parameter.
\section{Phase estimation}\label{s:phase}
The quantum-mechanical measurement of the phase of the radiation field
is the essential problem of high sensitive interferometry, and has
received much attention in quantum optics \cite{rev}.  The problem
arises because for a single mode of the electromagnetic field there is
no selfadjoint operator for the phase, hence a more general description
of the phase measurement is needed on the ground of estimation theory
\cite{hel,hol}.  \par In the following we 
apply the ML method to different schemes of phase estimation and
evaluate the corresponding sensitivity.
\subsection{Heterodyne detection on coherent state}
\par\noindent For a coherent state with amplitude $A\,e^{i\psi}$  
the probability density for 
complex outcome $\alpha _i$ at the $i$-th heterodyne measurement is given by
\begin{eqnarray}
p(\alpha _i)=\frac 1\pi \exp(-|\alpha _i -A\,e^{i\psi}|^2)\;.
\end{eqnarray}
The max-likelihood condition $\partial L/\partial \psi=0$ provides the
MLE for the phase $\psi $. One obtains $\psi _{\sc
ml}=\arg(\overline\alpha )$, where the overline denotes the
experimental average over N heterodyne outcomes, namely $\overline
\alpha=(\sum_{i=1}^N \alpha _i)/N$. For small phase-shift $\psi \simeq
0$ the Cram\'er-Rao bound gives the constraint $\sigma _\psi \geq
1/\sqrt{2nN}$, $n$ being the average photon number ($n=A^2$).
\subsection{Homodyne detection at  random phase on coherent state} 
\par\noindent In this case the homodyne probability for outcome $x_i$
at the $i$-th measurement at phase $\phi _i$ writes
\begin{eqnarray}
p(x_i,\phi_i)=\sqrt{\frac
2\pi}\exp\left\{-2[x_i-A\cos(\phi_i-\psi)]^2\right\}\;.
\end{eqnarray}
The ML condition provides for the estimator of $\psi $ the solution 
\begin{eqnarray}
\psi _{\sc ml}=\hbox{arctg}\left(\overline {x\sin \phi}/
\overline {x\cos\phi}\right)
\;.
\end{eqnarray}
Also in this kind of phase-detection strategy, the variance of the 
estimator for small phase-shift satisfies 
\begin{eqnarray}
\sigma_\psi^2  \geq
\frac {1}{2nN}
\;.\label{cr1}
\end{eqnarray}
\subsection{Homodyne detection at fixed phase on squeezed states}
\par
\begin{figure}[h]
\begin{center}
\psfig{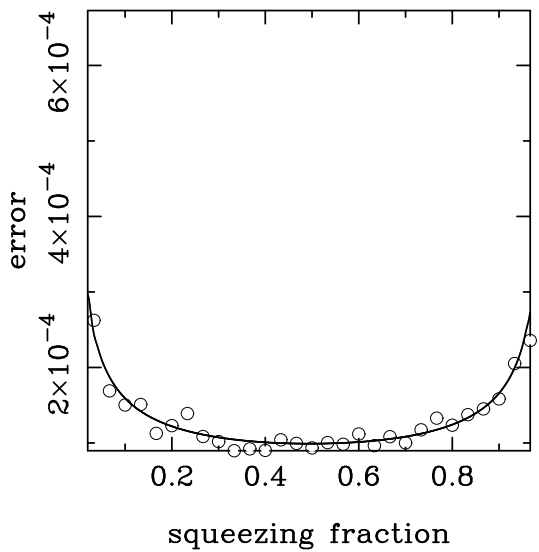}
\end{center} 
\caption{Estimation of the statistical error (circles) for the phase-shift
measurement through the maximum likelihood method on a squeezed state 
of radiation, for different values of the degree of
squeezing. The total number of photon of the state is fixed at
$n=50$. The solid line represents the Cram\'er-Rao bound on the
errors. Only 5000 homodyne data have been used, and the bound is
saturated, thus proving the efficiency of the method.}
\label{f:homo}
\end{figure}

\begin{figure}[h]
\begin{center}
\psfig{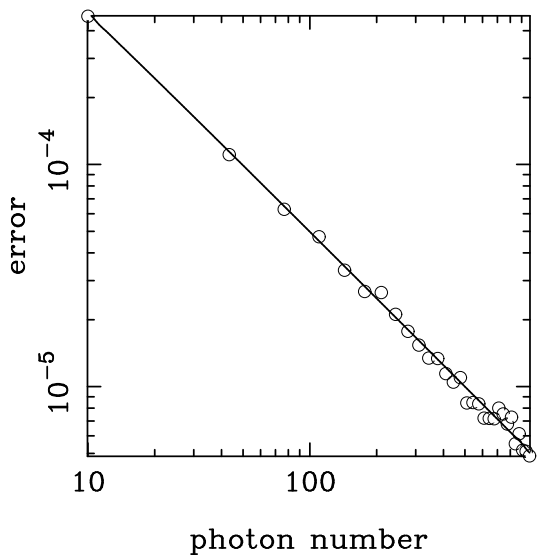}
\end{center} 
\caption{Phase sensitivity {\em vs} total number of photons  
achievable through homodyne detection on squeezed states and maximum
likelihood estimation, with optimal fraction of squeezing photons [see
Eq. (\ref{e2r})]. Compare the results of a Monte-Carlo simulation with
5000 homodyne outcomes (circles) with the theoretical behavior (solid
line).}
\label{f:homosun}
\end{figure}

\noindent
The use of squeezed states and homodyne detection at the phase
corresponding to the squeezed quadrature offer a better result in
terms of sensitivity. Consider the problem of estimating the phase
$\psi$ in the state $D(A\,e^{i\psi})S(r)|0 \rangle $ 
with $A,r >0$. 
The homodyne probability of  outcome $y_i$ 
for the measurement of the quadrature $Y=(a-a^\dag )/2i$ writes
\begin{eqnarray}
p(y_i)=\sqrt{\frac{2\,e^{2r}}{\pi}}\exp\left[-2\,e^{2r}\,(y_i-A\sin
\psi)^2\right]
\;.\label{py}
\end{eqnarray}
The MLE for $\psi $ is then given by $\psi _{\sc ml}=\hbox{arcsin}
(\overline y/A)$. For small phase shift $\psi \simeq 0$ 
the Cram\'er-Rao bound provides the relation
\begin{eqnarray}
\sigma^2 _\psi \geq \frac {1}{4N \,A^2\,e^{2r}}.\;\label{s2}
\end{eqnarray}
Upon maximizing the product $A^2\,e^{2r}$ versus the total number of
photons in the state $n=A^2+\sinh ^2r$, one obtains the optimal
squeezing 
\begin{eqnarray}
e^{2r}=2A^2\left[1+\sqrt{1+\frac {1}{4A^4}}\right] 
\;.\label{e2r}
\end{eqnarray}
Notice that for $A\gg 1$, Eq. (\ref{e2r}) requires that an
equal number of squeezing and coherent photons contributes to the
total average power in the radiation, namely $n_{coh} 
\simeq n_{sq}\simeq n /2$. In this case Eq. (\ref{s2})
rewrites
\begin{eqnarray}
\sigma^2_\psi\geq \frac {1}{4N \,n^2}\;,\label{1n2}
\end{eqnarray}
namely one obtains the ideal limit for the sensitivity of phase
estimation \cite{hel,hol}.
The bounds on sensitivity obtained in the previous examples are
saturated within a rather small number of data samples.  In
Fig. \ref{f:homo} we compare the experimental error obtained by a
Monte-Carlo simulation of homodyne detection on squeezed states using
5000 data samples with the theoretical bound of Eq. (\ref{s2}).  We
fixed the total number of photons at the value $n=50$, and varied the
squeezing fraction $n_{sq}/n$.  Notice how experimental and
theoretical data compare very well. We estimated the statistical
errors in Figs. 2-4 from the raw data by propagation of the errors on
the evaluation of $\overline y$, namely
\begin{eqnarray}
\sigma _\psi^2=\left(\frac{\partial \psi_{\sc ml}}{\partial \overline
y}\right)^2\,\sigma^2 _{\overline y} \;.\label{s22}
\end{eqnarray}
Notice that for large data samples, $\sigma _{\overline y} 
\rightarrow e^{-2r}/4N$, and one recovers Eq. (\ref{s2}). As shown in
Figs. 2-4, our estimation of errors approaches the Cram\'er-Rao bound,
hence proving that the ML method for the phase estimation is
statistically efficient. 
At the optimal value of squeezing fraction [see Eq. (\ref{e2r})], 
the behavior $\sigma _\psi\propto 1/n$ is well reproduced, also at
the small number 5000 of data samples, as shown in
Fig. \ref{f:homosun}.  
\par
\begin{figure}[h]
\begin{center}
\psfig{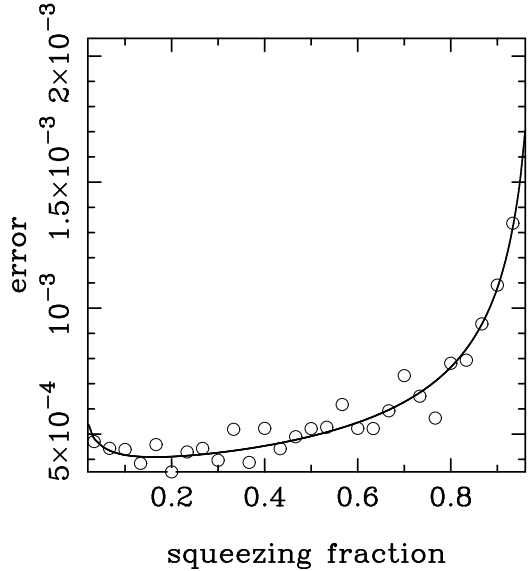}
\end{center} 
\caption{Same as in Fig. \ref{f:homo}, but for quantum efficiency
$\eta =80\%$. Notice how the best sensitivity is achieved for a
smaller fraction of squeezing photons.}\label{f:homoeta}
\end{figure}

 Unfortunately, the result in Eq. (\ref{1n2}) is very sensible to the
effect of less-than-unity quantum efficiency $\eta $ of realistic homodyne
detectors. For $\eta <1$, the homodyne probability is given by a
convolution of the ideal distribution in Eq. (\ref{py}) with a
Gaussian with  variance $(1-\eta )/4\eta$. In such case,
Eq. (\ref{s2}) is replaced by 
\begin{eqnarray}
\sigma^2_\psi\geq \frac {e^{-2r}+\frac
{1-\eta}{\eta}}{4N \,A^2}\;.\label{s2eta} 
\end{eqnarray}
The optimal value of the squeezing factor $e^{-2r}$ to minimize
Eq. (\ref{s2eta}) at fixed total number of photons is given by the
solution in the interval $I=[0,1]$ of the cubic equation
\begin{eqnarray}
x^3+ \left(4A^2+\frac{1-\eta}{\eta}\right)x^2-x -\frac {1-\eta
}{\eta}=0
\;.
\end{eqnarray} 
Compare Fig. \ref{f:homo} with Fig. \ref{f:homoeta}, where quantum
efficiency $\eta =0.8$ has been used. Indeed, the optimal squeezing
fraction rapidly approaches zero as $r\simeq A^2\,\eta $ for $\eta
\rightarrow 0$. Such
a detrimental effect of quantum efficiency is similar to the effect of
losses in squeezed-state homodyne communication channels \cite{qopt}. 
However, it can be partially stemmed by adopting a feedback-assisted 
homodyne detection \cite{fed}.
\section{Absolute estimation of the quantum efficiency}\label{s:eta}
In principle, in a photodetector each photon ionizes a single atom,
and the resulting charge is amplified to produce a measurable
pulse. In practice, however, available photodetectors are usually
characterized by a quantum efficiency lower than unity, which means
that only a fraction of the incoming photons lead to an electric
pulse, and ultimately to a "count". If the resulting current is
proportional to the incoming photon flux we have a linear
photodetector. This is , for example, the case of the high flux
photodetectors used in homodyne detection.  On the other hand,
photodetectors operating at very low intensities resort to avalanche
process in order to transform a single ionization event into a
recordable pulse. This implies that one cannot 
discriminate between a single photon or many photons as the 
outcomes from such detectors are either a "click",
corresponding to any number of photons, or "nothing" which means that
no photons have been revealed. In this section we apply the ML
principle to the absolute estimation of the quantum efficiency of both
linear and avalanche photodetectors. We suppose to have at our 
disposal a known reference state and, from the results of a
measurement upon such a state, we infer the value of the
quantum efficiency. \par
\begin{figure}[h]
\begin{tabular}{c}
\psfig{file=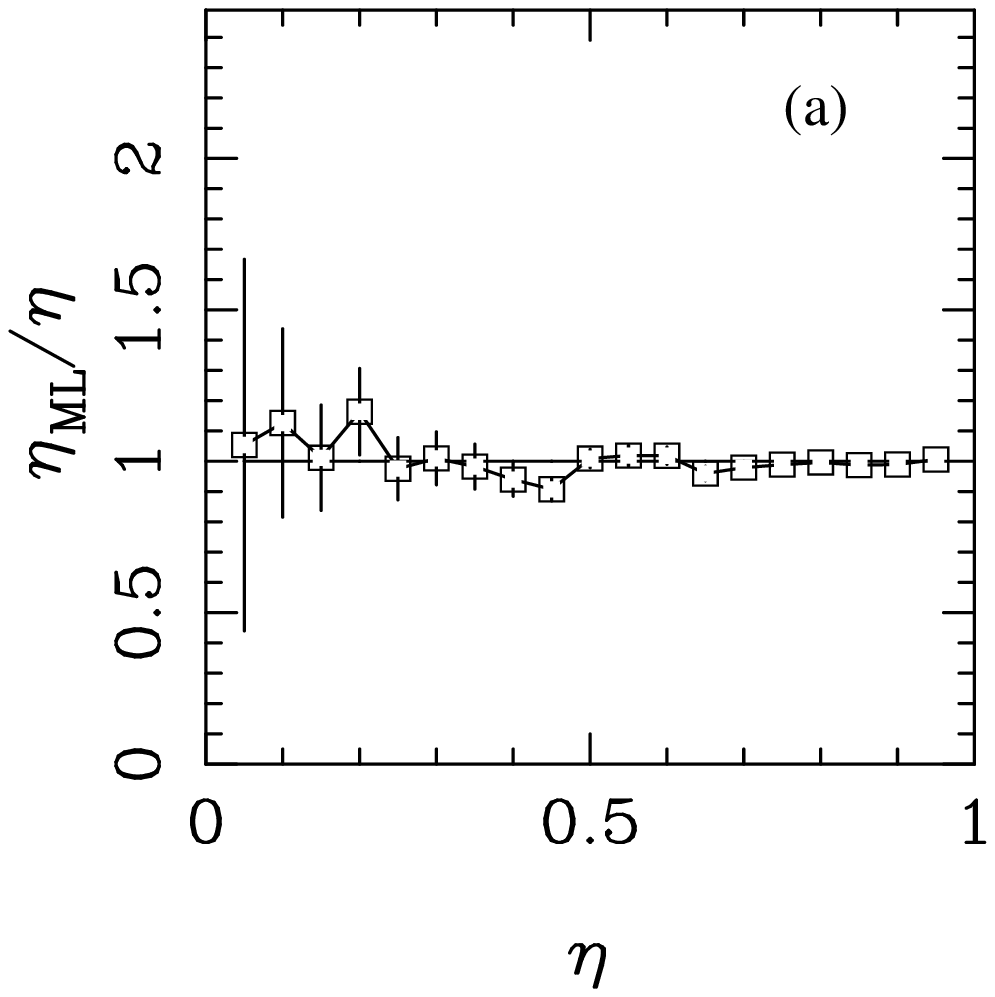,width=7cm} \\
\psfig{file=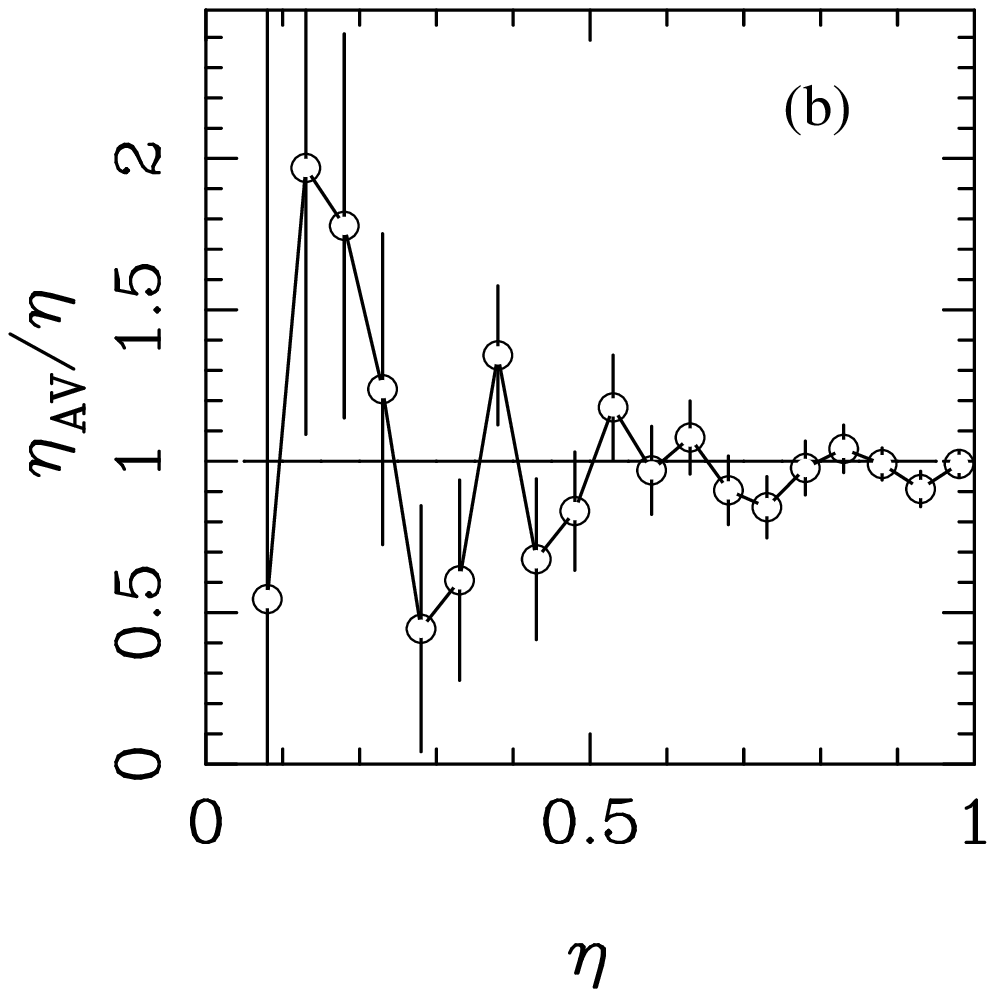,width=7cm} 
\end{tabular}
\caption{Estimation of the quantum efficiency of linear photodetectors
through homodyne detection on a squeezed state. Both plots report the
ratio between the estimated value of the quantum efficiency and the
true value, as a function of the true value. 
(a) results obtained using the maximum-likelihood method; (b) 
results by the "naive" average-value method. The homodyne
sample consists of $50$ blocks of $50$ data each, whereas the
reference state is a squeezed state with mean number of photons $n=1$
and squeezing fraction $\gamma = 99 \%$ (nearly a squeezed vacuum).}
\label{f:eta}
\end{figure}

 Let us first study the case of linear
photodetectors. As a reference state we consider a squeezed-coherent
state, measured by homodyne detection.  The effect of  nonunit
quantum efficiency $\eta$ on the probability distribution of homodyne
detection is twofold. We have both a rescaling of the mean value and a
broadening of the distribution.  For a squeezed state $|x_0,r\rangle =
D (x_0) S(r) |0\rangle$ with the direction of squeezing
parallel to the signal phase and to the phase of the homodyne
detection (without loss of generality we set this phase equal to zero
and $x_0,r>0$ ) we have \cite{paul}
\begin{eqnarray}
p_\eta (x) &=& \frac1{\sqrt{2\pi\Delta^2}} 
\exp\left[-\frac{(x-\eta x_0)^2}{2\Delta^2}\right]\;,\nonumber \\
\Delta^2 &=& \frac14  \left(e^{-2r}+1-\eta\right)\label{pheta}\;.
\end{eqnarray}
The total number of photons of the state is given by $ n=x_0^2+\sinh^2 r $, 
whereas the squeezing fraction is
defined as $\gamma = \sinh^2r /n$.  Apart from an irrelevant constant, the 
log-likelihood function can be written as
\begin{eqnarray}
- L(\eta )= \log \Delta^2 + \frac{1}{\Delta^2}\left(\overline{x^2}+\eta x_0^2 -
2 \eta x_0 \overline{x}\right)
\label{loglikH}\;.
\end{eqnarray}
The resulting MLE is thus given by
\widetext
\begin{eqnarray}
\eta_{\sc ml} = 1+ e^{-2r} + \frac1{x_0^2} \left\{
1-\sqrt{1+64x_0^2\left[\overline{x^2}+(1+e^{-2r})(x_0-2\overline{x}
+x_0e^{-2r})x_0\right]}\right\}
\label{etaMLE}\;.
\end{eqnarray}
\narrowtext
A set of Monte Carlo simulated
experiments confirmed that the Cram\'er-Rao bound is attained.  The
performances of the ML estimation can be compared to the "naive"
estimation based only on the measurement of the mean value, i.e.
$\eta_{\sc av} = \overline{x}/x_0$. We expect this method to be less
efficient, since the quantum efficiency not only rescales the mean
value, but also spreads the variance of the homodyne distribution in
Eq. (\ref{pheta}).  In Fig. \ref{f:eta}, on the basis of a Monte Carlo
simulated experiment, we compare the ML and the average-value methods
in estimating the quantum efficiency through homodyne detection on a
squeezed state.  The advantages of ML method are apparent, especially
for the estimation of low values of $\eta$. On the other hand, for
small values of the squeezing fraction the two methods have similar
performances, except for very low signals, whereas the ML estimation
performs better. \par Let us now consider avalanche photodetectors,
which perform the {\sf ON/OFF} measurement described by the two-value
POM
\begin{eqnarray}
\Pi_{\sc off} =  \sum_{p=0}^{\infty} (1-\eta)^p \: |p \rangle\langle p|
\qquad  \Pi_{\sc on} = {\bf I} - \Pi_{\sc off}
\label{yesno}\;, 
\end{eqnarray}
where ${\bf I}$ denotes the identity operator. Indeed, for
high quantum efficiency (close to unity) $\Pi_{\sc off}$ and
$\Pi_{\sc on}$ approach the projection operator onto the vacuum
state and its orthogonal subspace, respectively. With avalanche
photodetectors we have only two possible outcomes: "click" or "no
clicks" which we denote by "1" and "0" respectively. The
log-likelihood function is given by
\begin{eqnarray}
L(\eta) = (N - N_c) \log P_0 (\eta) + N_c \log [1-P_0(\eta)]
\label{loglikyn}\;,
\end{eqnarray}
where $P_0 (\eta)=\hbox{Tr} [\varrho  \Pi_{\sc off}]$ is the
probability of having no clicks for the reference state described by
the density matrix $\varrho$, $N$ is the total number of
measurements, and $N_c$ is the number of events leading to a click.  The
maximum of $L(\eta)$, i.e. the MLE for the quantum efficiency,
satisfies the equation
\begin{eqnarray}
P_0 (\eta_{\sc ml}) = 1 - \frac{N_c}{N}
\label{MLyn}\;,
\end{eqnarray}
whose solution, of course, depends on the choice of the reference state.  The
optimal choice would be using single-photon states as a
reference. In this case, we have the trivial result $\eta_{\sc ml}=
N_c/N$. However, single-photon state are not easy to prepare \cite{fock}, 
and generally one would like to test $\eta $ for coherent pulses
$|\alpha\rangle$. In this case, we have $P_0(\eta)= \exp (-|\alpha
|^2\eta)$ and
\begin{eqnarray}
\eta_{\sc ml} = - \frac1{|\alpha |^2} \log \left( 1 - \frac{N_c}{N}\right)
\label{MLEyn}\;.
\end{eqnarray}
The Fisher information is given by
\begin{eqnarray}
F &=& \left(\frac{\partial P_0 }{\partial\eta}\right)^2 \frac1{P_0} + \left(
\frac{\partial P_1}{\partial\eta}\right)^2 \frac1{P_1} \nonumber \\ 
&=& \frac1{P_0(1-P_0)}\left(\frac{\partial P_0 }{\partial\eta}\right)^2 
\label{Fisheryn}\;,
\end{eqnarray}
and therefore, for a weak coherent reference one has  
\begin{eqnarray}
F = \frac{\eta^2}{e^{\eta |\alpha|^2}-1}\simeq \frac\eta{|\alpha|^2} 
\label{confyn}\,
\end{eqnarray}
and 
\begin{eqnarray}
\sigma_\eta &\simeq& \frac{|\alpha|}{\sqrt{\eta N}}
\;.\label{}
\end{eqnarray}
\section{Summary and conclusions}\label{s:outro}
In quantum optics, there are several parameters of great interest
corresponding to quantities that are not directly observable. Among
these, we studied the parameters of a Gaussian state, the phase of a
squeezed-coherent state, and the quantum efficiency of either linear
or single-photon resolving photodetector.  In this paper, we have
applied the maximum-likelihood method to the determination of these
parameters using feasible detection schemes. In particular, we have
considered homodyne detection and {\sc on/off} photodetection.  In all
cases here analyzed, the resulting estimators are efficient, unbiased
and consistent, thus providing a statistically reliable determination
of the parameters of interest. Moreover, by using the ML method only
few thousand data are required for the precise determination of
parameters. We stress that the ML procedure used in this paper can be
applied to a broad class of estimation process, since it applies to
any probability distribution $p(x|\lambda)$, as long as its functional
form is known and the maximum of the likelihood function is unique. In
conclusion, for the measurement of parameters pertaining
to quantum states or optical devices, the ML procedure should be
taken into account, in order to optimize data analysis and thus
reducing the experimental efforts.

\twocolumn
\end{document}